# Hurricane recovery gaps lie in reconnection, not local restart

Angelina Liu[1], Mengqiu Cao[2], Sen Pei[3], Qing Yao[3*]

[1] Scarsdale High School, Scarsdale, NY, USA

[2] Bartlett School of Environment, Energy and Resources, University College London, London, UK

[3] Environmental Health Sciences, Columbia University Mailman School of Public Health, New York, USA

[*] Corresponding author: qy2290@cumc.columbia.edu

**Abstract**

Climate-intensified hurricanes disrupt urban mobility, yet whether recovery is equitable across the rural–urban gradient is unclear. Analyzing device-based mobility for Hurricanes Helene and Milton, we find local activity restarted within ~4.5 days in both storms, whereas reconnection to the outside world took 5 days in coastal-urban Florida but 11 in inland-rural Appalachia. The gap lies in external reconnection, not local restart, implying planners should target rural isolation.

**Main text**

Tropical cyclones are intensifying under climate change[1], and their public-health and economic burdens are rising accordingly[2–4]. Human mobility, the daily movement that sustains work, care, schooling and commerce offers a sensitive readout of how urban systems absorb and recover from these shocks[5]. Most disaster-mobility research has nonetheless concentrated on evacuation: who leaves before landfall, and how that decision is stratified by income, race and vehicle access[6–9]. Recovery has received less attention, and where it has been studied it is typically for a single event in a single metropolitan region[10–12], leaving unresolved whether communities differing in urbanicity recover at comparable speed. This matters because the equity-relevant question for adaptation planning is not only who can leave, but who is left waiting. The 2024 Atlantic season offers an unusual natural comparison: Hurricanes Helene (landfall 26 September 2024) and Milton (9 October 2024) struck the southeastern United States 13 days apart, one inland and rural, the other coastal and urban. Prior work compared their aggregate regional mobility during the landfall week[13] but did not measure recovery, resolve activity categories, or examine local spatial structure. Here we decompose mobility into within-area, inflow and outflow components across six activity categories and two spatial scales and show that the rural–urban recovery gap lies almost entirely in external reconnection rather than in local restart. We analyzed counties within 100 miles of each storm's IBTrACS track[14], 487 counties for Helene and 34 for Milton, together encompassing 42.9 million residents. This boundary captures Helene's inland flooding footprint, including southern Appalachia.

Milton delivered the deeper shock, and every activity category agrees. Fig. 1 shows daily deviation from the counterfactual baseline for each of six categories, from one week before landfall to two weeks after. Within-area mobility fell by 24–45% across categories under Milton against 8–13% under Helene, between two and four times deeper in every category, and inflow fell by 27–60% against 20–27%. The outflow in Milton affected region surged 2 days before the landfall while Helene affected region exhibited shallow increase and decrease before and after landfall. Disruption varied across activities. Under Milton, Travel showed the largest declines in both within-area mobility (45%) and inflow (60%), whereas Health visits fell less sharply (24% and 27%, respectively).

The storms differed in how long disruption persisted as well as how sharply mobility declined. Milton's within-area decline was deeper but brief, whereas Helene's inflow deficit continued well beyond the landfall week. Recovery estimates show this distinction more clearly (Fig. 1g,h). Within-area mobility recovered in 3.2–5.1 days under Helene and 4.7–6.1 days under Milton. Helene recovered sooner in five of six categories, with the largest difference in Education: 3.2 days compared with 6.1 days after Milton.

For incoming visits, however, Helene consistently recovered more slowly. Across all six categories, inflow took 11.3–13.3 days to return to baseline after Helene, compared with 4.4–5.5 days after Milton. This delay extended beyond Travel to visits associated with work, healthcare and education. Incoming Health visits, for example, took 12.1 days to recover after Helene, compared with 4.6 days after Milton. The longer inflow recovery therefore affected several parts of daily life, even as movement within the affected region had largely recovered.

The regional results show how recovery differed between storms, but they can hide substantial differences between local areas within each affected region. We therefore examined local mobility to identify where disruption was greatest and whether it was associated with storm exposure, socioeconomic conditions or settlement patterns. For local analysis, we pooled visits across categories and grouped Helene's counties into 101 contiguous clusters sharing an urban–rural classification[15] to stabilize low-volume mobility baselines; Milton's 34 counties were analyzed individually. A flow-specific threshold of 20,000 visits per day retained fewer Helene clusters and Milton counties different flow types (Methods and Supplementary Information). The same pattern of visits to all categories appears when we examine individual counties for Milton and groups of neighboring counties for Helene. Mobility within these areas recovered at similar rates: median recovery time was 4.4 days for Helene (IQR 3.7–5.1; n = 96) and 4.5 days for Milton (IQR 4.1–5.4; n = 33), with no statistically significant difference between storms (Mann–Whitney P = 0.294). In contrast, inflow recovery took more than twice as long after Helene (median 11.1 days, IQR 8.7–13.4; n = 46 clusters) as after Milton (median 5.1 days, IQR 4.2–6.2; n = 28 counties; Mann–Whitney P < 0.001; Fig. S1). The longer recovery after Helene therefore reflects a slower return of incoming travel, even as movement within local areas resumed.

Fig. 2a shows where mobility declined most sharply in each affected region. Under Helene, the largest inflow declines (35–46% below baseline) occurred farther inland while under Milton, the regions closer to landing exhibit larger drops or higher increase. Is storm exposure the only factor for where mobility disruption was greatest? We fitted separate models for Helene and Milton, allowing associations between mobility disruption, storm exposure and local characteristics to differ between events (Fig. 2b). Under Helene, coastal areas experienced deeper within-area declines, heavier rainfall was associated with larger inflow declines, and outflow increases were smaller farther from the storm track. A higher proportion of White residents was also associated with deeper inflow declines. These four associations had 95% HDIs excluding zero. Milton's estimates were less conclusive: all intervals included zero, with uncertainty remaining substantial in its smaller sample.

The percentage-White association, however, weakened to near zero after accounting for broad spatial patterns in Helene's inflow disruption, while the coastal association became stronger (Fig. 2c). Adding geographic adjustment increased deviance explained from 15% to 52%. This suggests that population composition and geography share explanatory information, making their contributions difficult to distinguish. The result does not establish that socioeconomic conditions are unimportant, but it limits how confidently we can attribute the observed disruption to them.

These results reframe what post-hurricane resource allocation should target. Milton produced deeper mobility declines, but incoming travel recovered sooner than after Helene. This comparison shows why the initial decline alone gives an incomplete picture of recovery. Places with the largest initial mobility declines were not necessarily those with the longest recovery. Although movement within local areas recovered at similar rates after both storms, incoming travel took substantially longer to recover after Helene. The depth of the initial decline alone may therefore miss communities facing prolonged disruption. Recovery planning could also track how quickly incoming travel returns, helping identify places where local movement has resumed but connections with the wider region remain disrupted.

This study contributed the measurement of post-hurricane recovery by distinguishing the resumption of local activity from the return of incoming movement. Combining counterfactual mobility baselines with flow decomposition across six activity categories and two spatial scales reveals a recovery burden that initial disruption alone can miss. Despite Milton's deeper mobility declines, inflow recovery took more than twice as long after Helene as after Milton across all six regional activity categories. Local activity can rebound while incoming travel remains depressed, leaving an incomplete picture of recovery. Incorporating both into disaster monitoring could help emergency managers flag communities for checks of transport links and essential-service access. Combined with these assessments, persistent inflow deficits could inform where recovery support remains needed.

Three limitations qualify these findings. First, Helene and Milton affected regions with different storm conditions, geography and settlement patterns. With only two events, we cannot determine which of these differences explains Helene's longer inflow recovery. Comparisons across more storms using consistent spatial units are needed to assess whether the recovery contrast holds more broadly. Second, device-based visits do not identify trip purpose and may under-represent rural residents, older adults and people with low incomes[5]. Coverage gaps and measurement disruption may affect recovery estimates in an uncertain direction, while the volume threshold limits generalization to low-volume communities. Finally, our recovery measure estimates when visit counts return to expected levels; it does not establish whether that return persists or whether essential services have fully recovered.

## Methods

**Mobility data.** Daily origin–destination visit counts derive from device-based foot-traffic records (Advan Research Weekly Patterns, distributed via Dewey Data[16]). Point-of-interest (POIs) categories were aggregated into six activity groups: Travel, Work & Professional, Health, Education, Retail & Leisure, and Urban Government (Table. S1). For each unit which can be the whole affected regions or the local areas, we form three daily flows: within-area visits had both endpoints inside the focal unit; inflow originated outside the affected region and terminated inside the unit; outflow followed the reverse direction.

**Counterfactual baseline.** For each spatial unit and flow, we fitted a calendar regression with AR(1) errors to log-transformed daily visits $M(t)$:

$$log(1 + M(t)) = c + \sum_{d=1}^{6} \beta_d D_d(t) + \sum_{m=8}^{10} \gamma_m H_m(t) + \delta Y(t) + \eta(t),$$
$$\eta(t) = \varphi\eta(t - 1) + \varepsilon(t). \quad (1)$$

Here, $D_d(t)$, $H_m(t)$ and $Y(t)$ indicate weekday, month and 2024, respectively, with Monday, July and 2023 as reference categories. Models were fitted by state-space maximum likelihood using July–October 2023 and pre-event 2024 observations. The 2024 training cutoffs were 19

September for Helene, 2 October for Milton's regional series and 31 August for Milton's local series, the latter to avoid contamination from Helene. If $\hat{z}(t)$ denotes the predicted log volume, the counterfactual is $\hat{M}(t) = exp(\hat{z}(t)) - 1$. Disruption is measured as the relative deviation $rd(t) = 100(\frac{M(t) - \hat{M}(t)}{\hat{M}(t)})$.

**Metrics.** Largest drop was the minimum $rd(t)$ from landfall through day 6 for within-area and inflow mobility. Outflow surge was the maximum relative deviation from three days before landfall through day 6. Recovery used a centered three-day moving average. We located the trough ($t_T$) within 10 days after landfall, then fitted a Theil–Sen line[17,18] to its nondecreasing recovery segment, ending at the first decline. Recovery time was:

$$\tau = (t_T - t_L) - \frac{a}{b}.$$

where, $t_L$ is the landfall and $a$ and $b$ are the fitted intercept and slope. Estimates required a negative trough, at least two observations and a positive slope. The fitted return to baseline may extend beyond the observed segment. Outflow recovery was not estimated.

**Sample size restriction.** Per flow, units with mean observed volume below 20,000 visits/day were excluded, removing a noisy low-volume tail (Table. S2). Recovery distributions were compared using two-sided Mann–Whitney tests.

**Regression.** Separate regularized Bayesian models related each storm's disruption magnitudes to nine predictors (Table. S3), including 7-day precipitation[19], distance to track, coastal indicator, urban–rural classification code[15], median income[20], % White[20], insurance coverage[20], % without vehicle[20], and population density[20]. Continuous predictors were standardized; coastal status remained binary. Residual spatial dependence was assessed by Moran's I under Queen contiguity[21]. A generalized additive model with a penalized spatial smooth assessed sensitivity to broad geographic patterns[22]. Pooled models assessing sensitivity to the affected-region boundary are described in the Supplementary Information.

## References

1. IPCC. Climate Change 2022: Impacts, Adaptation and Vulnerability (2023). doi:10.1017/9781009325844
2. Parks, R. M. & Guinto, R. R. Tropical cyclones and public health. doi:10.1289/EHP12241
3. Parks, R. M. et al. Short-term excess mortality after tropical cyclones. Sci. Adv. (2023). doi:10.1126/sciadv.adg6633
4. NOAA NCEI. U.S. Billion-Dollar Weather and Climate Disasters.
5. Yabe, T. et al. Mobile phone location data for disasters. Comput. Environ. Urban Syst. (2022). doi:10.1016/j.compenvurbsys.2022.101777
6. Thompson, R. R., Garfin, D. R. & Silver, R. C. Evacuation from natural disasters. Risk Anal. (2017). doi:10.1111/risa.12654
7. Deng, H. et al. Race and wealth disparities in evacuation (2021). doi:10.1057/s41599-021-00824-8
8. Huang, S.-K., Lindell, M. K. & Prater, C. S. Hurricane evacuation meta-analysis (2016). doi:10.1177/0013916515578485
9. Anand, H. et al. Income and race disparities in hurricane evacuation are contingent upon study case and design. Sci. Rep.
10. Wang, Q. & Taylor, J. E. Hurricane Sandy mobility resilience. PLoS ONE (2014). doi:10.1371/journal.pone.0112608
11. Wang, S. et al. Hurricane Ian mobility vulnerability and recovery (2024). doi:10.1080/21681376.2024.2369550
12. Hong, B. et al. Inequality in community resilience from mobility data. Nat. Commun. (2021). doi:10.1038/s41467-021-22160-w
13. Yao, Q. et al. Adaptive mobility responses during Hurricanes Helene and Milton in 2024. Environ. Res. Lett. (2025). doi:10.1088/1748-9326/ae0e39
14. NOAA IBTrACS. International Best Track Archive for Climate Stewardship.

15. National Center for Health Statistics. NCHS Urban–Rural Classification Scheme for Counties, 2023 Scheme (National Center for Health Statistics, 2023); https://www.cdc.gov/nchs/data-analysis-tools/urban-rural.html
16. Advan Research, Weekly Patterns, via Dewey Data. doi:10.82551/X1PP-1F65 [VERIFY]
17. Theil, H. A rank-invariant method of linear and polynomial regression analysis (1950).
18. Sen, P. K. Estimates of the regression coefficient based on Kendall's tau. JASA (1968). doi:10.1080/01621459.1968.10480934
19. PRISM Climate Group. PRISM Time Series Data (Oregon State University, 2024); https://prism.oregonstate.edu/data/
20. U.S. Census Bureau. American Community Survey 2022 5-year estimates.
21. Moran, P. A. P. Notes on continuous stochastic phenomena. Biometrika (1950). doi:10.1093/biomet/37.1-2.17
22. Wood, S. N. Generalized Additive Models: An Introduction with R 2nd edn (CRC Press, 2017).
23. Hodges, J. S. & Reich, B. J. Adding spatially-correlated errors can mess up the fixed effect you love. Am. Stat. (2010).
24. Paciorek, C. J. The importance of scale for spatial-confounding bias. Stat. Sci. (2010).

## Figures

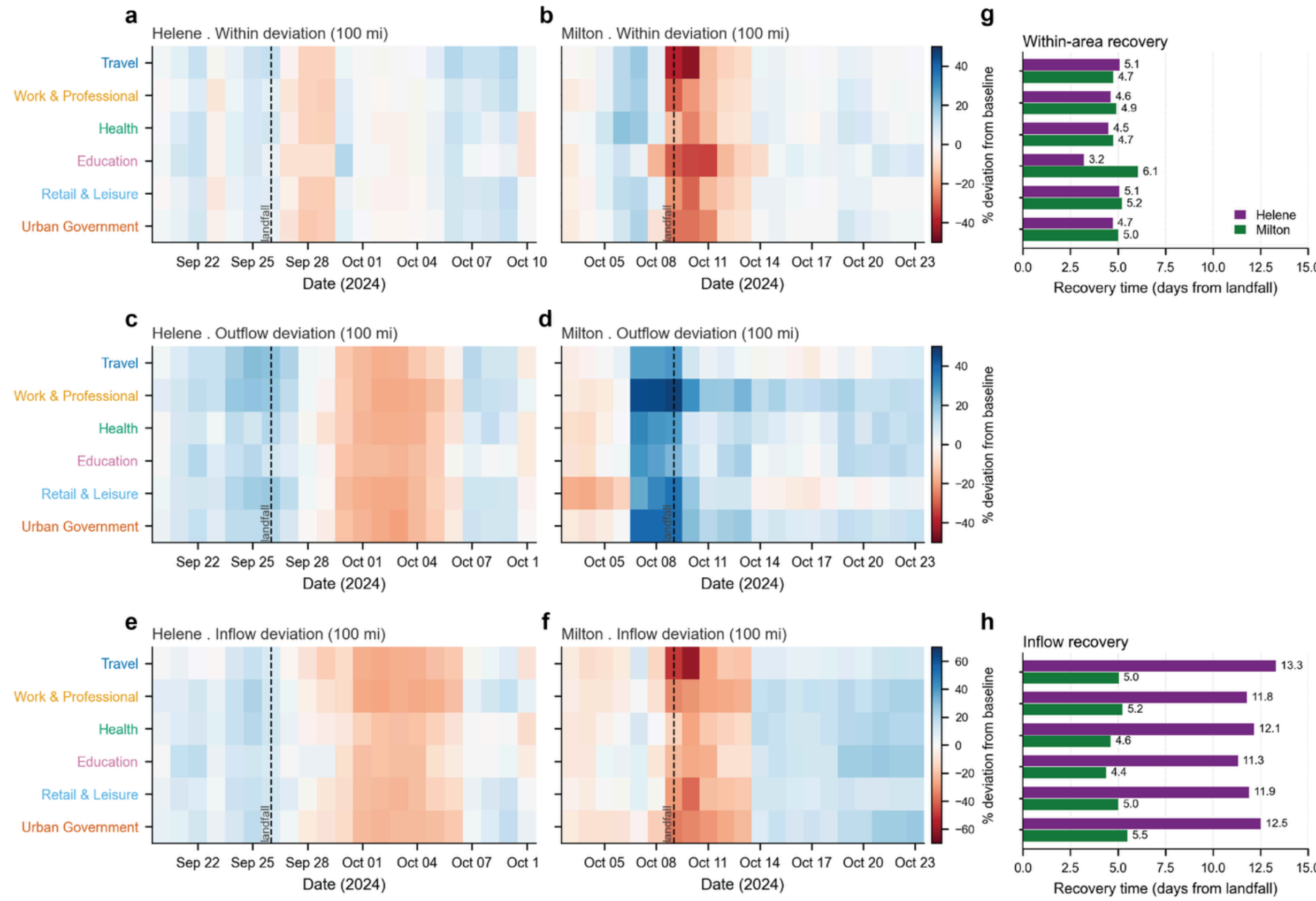

**Fig. 1 Daily mobility-deviation heatmaps and recovery-time bar plots for Hurricanes Helene and Milton.** Regional mobility by activity category, six categories × 22 days (landfall −7 to +14). Daily within-area deviation from the counterfactual baseline for Helene **(a)** and Milton **(b)**. Daily outflow deviation for Helene **(c)** and Milton **(d)**. Daily inflow deviation for Helene **(e)** and Milton **(f)**. Red denotes a decrease relative to baseline and blue a surplus; each flow uses a colour scale shared between storms, and the inflow scale is wider than the other two. **(g)** Estimated within-area recovery time, days from landfall until the category series is estimated to return to baseline (Theil–Sen zero crossing; Methods) — Helene (purple) against Milton (green). The same for inflow in **(h)**. Milton's within-area shock is roughly two- to four-fold deeper (b

versus a), yet only inflow recovery separates the storms (h); within-area recovery does not (g).

Outflow is a surge rather than a decline and therefore has no recovery time.

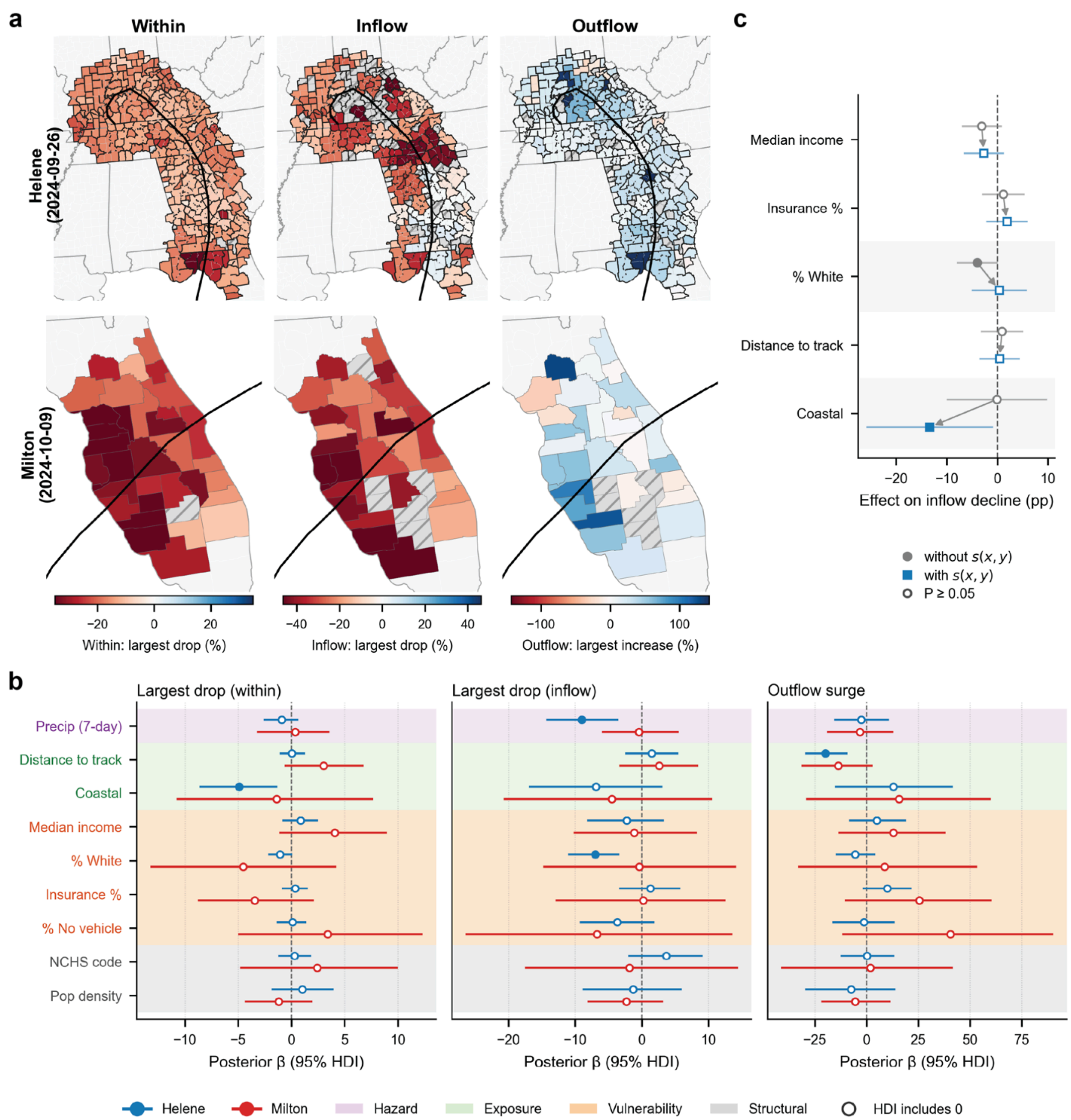

**Fig. 2 Maps of peak mobility deviation and regression coefficients for nine predictors, Hurricanes Helene and Milton. (a),** Largest per-unit deviation for each of the three flows — within-area (left), inflow (center) and outflow (right) — for Helene (top row, 101 clusters) and Milton (bottom row, 34 counties). Color scales are shared between storms within a flow but differ between flows; the black line is the IBTrACS best track. Hatched units are units with no

valid estimate. **(b)**, Posterior means and 95% highest-density intervals from per-storm regularized Bayesian regressions of three flow outcomes on nine standardized predictors (continuous predictors z-scored; coastal status retained as a binary contrast); Helene in blue and Milton in red, filled markers denote intervals. Sample sizes are n = 96, 52 and 74 units for Helene and n = 33, 28 and 29 for Milton, for within-area, inflow and outflow respectively. **(c)**, Helene inflow coefficients before (grey circles) and after (blue squares) adding a penalized two-dimensional thin-plate smooth s(x,y) to the regression; open markers denote $P \geq 0.05$ and filled markers $P < 0.05$, from two-sided Wald tests in a generalized additive model, n = 52 clusters. The percentage-White association attenuates to zero while coastal location strengthens; the smooth itself is significant ($P = 0.023$) and raises deviance explained from 0.15 to 0.52.

# Supplementary Information for

# Hurricane recovery gaps lie in reconnection, not local restart

Angelina Liu[1], Mengqiu Cao[2], Sen Pei[3], Qing Yao[3*],

1 Scarsdale High School, Scarsdale, NY, 10583, USA

2 Bartlett School of Environment, Energy and Resources, University College London, London, UK

3 Environmental Health Sciences, Columbia University Mailman School of Public Health, New York, USA

* Corresponding author: qy2290@cumc.columbia.edu

### Study geography and mobility flows

The affected region for each storm comprises counties whose centroids lie within 100 miles of the best track (Fig. S1). Helene made landfall on 26 September 2024 and Milton on 9 October 2024. The selected regions contain 487 and 34 counties, respectively. Helene counties are joined into 101 contiguous clusters with a common NCHS urban–rural code; Milton retains 34 county units. Queen adjacency permits counties to share either an edge or a vertex. The 2023 NCHS classification has six ordered categories, with larger codes indicating more rural settings, see Fig. S2.

For a local unit U within affected region R and daily origin–destination visits V, within-area flow sums visits with both endpoints in U. Inflow sums visits originating outside R and terminating in U; outflow sums visits originating in U and terminating outside R. Travel between U and another unit inside R is therefore not external inflow or outflow. At regional scale, U is replaced by R. These are visit counts, not counts of unique people or direct measurements of infrastructure service.

### Additional implementation details for counterfactual mobility and recovery estimation

Counterfactual models were implemented using SARIMAX with order (1, 0, 0), a constant and no stochastic seasonal component. Log-scale predictions were back-transformed directly without a lognormal mean correction; zero counterfactual predictions were treated as missing denominators. The centered three-observation moving average allowed shorter windows at series endpoints. Outflow maxima could remain negative when mobility stayed below its counterfactual throughout the evaluation window. Units without valid recovery estimates were excluded from recovery summaries. Forecast and fitted-slope uncertainty were not propagated into the reported recovery summaries.

A Theil–Sen line was fitted to the nondecreasing recovery segment. This robust linear estimator calculates the slope as the median of slopes between all pairs of observations, reducing sensitivity to outliers. The fitted line's zero crossing gives the estimated time at which mobility returns to its counterfactual baseline.

**Spatial adjustment**

To assess whether estimated covariate associations depended on broad geographic patterns, we fitted two generalized additive models: a nonspatial model (M0), containing five linear covariate terms, and a spatial model (M1), containing the same terms plus a penalized two-dimensional geographic smooth. The model without spatial adjustment was:

$$Y_{if} = \alpha_f^{(0)} + \sum_{j=1}^{5} \beta_{jf}^{(0)} Z_{ijf} + \varepsilon_{if}^{(0)} \quad \text{(S1)}$$

The spatially adjusted model added a smooth geographic term:

$$Y_{if} = \alpha_f^{(1)} + \sum_{j=1}^{5} \beta_{jf}^{(1)} Z_{ijf} + g_f(x_i, y_i) + \varepsilon_{if}^{(1)} \quad \text{(S2)}$$

Here, $i$ indexes clusters and f indexes within-area, inflow or outflow mobility. $Y_{if}$ is the minimum unsmoothed percentage deviation for within-area and inflow mobility, or the maximum percentage deviation for outflow, using the evaluation windows defined in the main Methods. $Z_{ijf}$ contains median household income, insurance coverage, percentage White, distance to the storm track and coastal status. The four numeric covariates were standardized to mean zero and standard deviation one within each flow-specific Helene sample; coastal status remained 0/1. The intercept α and covariate coefficients β were estimated separately in M0 and M1. The error term ε was assumed independent and Gaussian with mean zero in each model.

The function $g_f(x_i, y_i)$ is a penalized two-dimensional thin-plate regression spline evaluated at each cluster's centroid. Coordinates were expressed in kilometres in the projected EPSG:5070 coordinate system. The smooth was centred to have mean zero over the fitted sample, separating it from the intercept. Models were fitted in R using mgcv, with a spatial basis dimension of 20 and smoothing selected by restricted maximum likelihood (REML). The penalty limits excessive variation in the fitted geographic surface. [1,6]

Comparing the covariate coefficients between M0 and M1 assesses how the associations change after allowing for broad spatial variation in mobility disruption. This sensitivity analysis uses five covariates rather than the nine predictors in the primary Bayesian models; its coefficients and P values therefore represent a separate model comparison.

For Helene inflow, the spatial smooth has P = 0.023 and deviance explained increases from 0.15 to 0.52. Table S7 reports selected coefficients before and after spatial adjustment. Smooth-term P values for within-area and outflow models are 0.87 and 0.86, respectively. These are GAM diagnostics, distinct from posterior HDIs in the Bayesian models. [1,6]

## Pooled sensitivity

We pooled eligible local units from Helene and Milton to assess associations across the two events, using the original percentage deviations in mobility as outcomes. Separate Gaussian generalized additive models were fitted for within-area decline ($n = 129$), inflow decline ($n = 80$) and outflow surge ($n = 103$). Each model included a hurricane indicator and nine covariates: precipitation, distance to the storm track, coastal location, median income, insurance coverage, White population share, households without a vehicle, urban–rural classification and population density. Continuous covariates were standardized within each storm and outcome sample, while coastal location remained binary (Table. S6). We compared models without spatial adjustment with models containing separate spatial smooths for each hurricane (basis dimension $k = 10$), fitted by restricted maximum likelihood. Covariate slopes were shared across storms. Spatial-model 95% confidence intervals incorporated smoothing-parameter uncertainty.

Spatial adjustment increased deviance explained from 50.0% to 64.0% for within-area decline, from 44.6% to 63.5% for inflow decline and from 19.9% to 30.2% for outflow surge. In the spatial inflow model, a one-standard-deviation increase in income was associated with a 5.28-percentage-point deeper decline ($\beta = -5.28$, 95% CI −10.13 to −0.44). The estimates for White population share ($\beta = 3.74$, −2.23 to 9.72), rainfall ($\beta = -3.54$, −7.13 to 0.04) and coastal location ($\beta = -6.70$, −15.27 to 1.87) remained uncertain. For outflow surge, greater distance from the track was associated with a smaller increase ($\beta = -18.67$, −27.52 to −9.83), while higher insurance coverage was associated with a larger increase ($\beta = 9.56$, 1.64 to 17.48), see Fig. S4

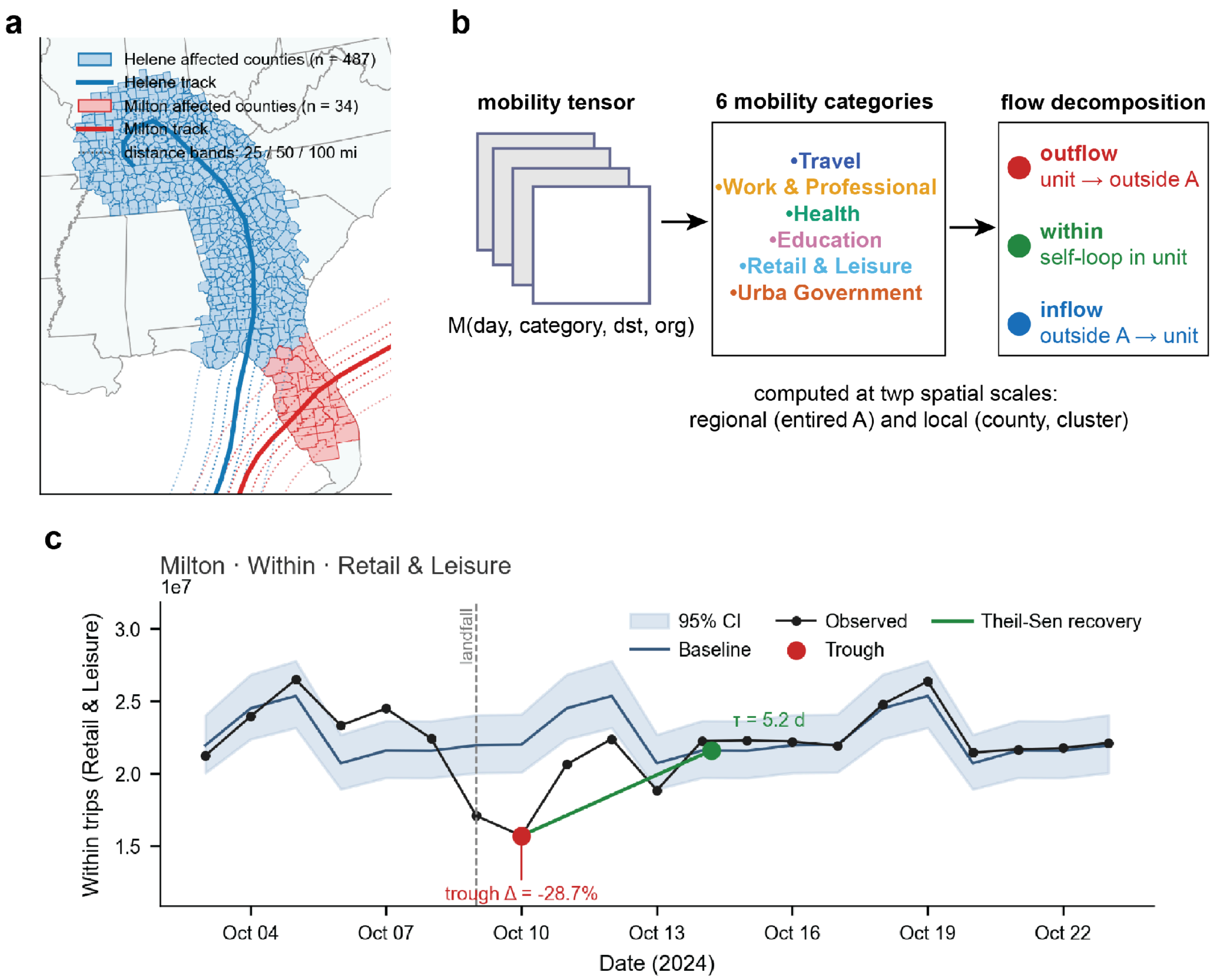

**Fig.S1 Study areas, mobility decomposition and recovery estimation. a.** Hurricane Helene and Milton tracks and affected counties, defined by the 100-mile track-distance cutoff (Helene, 487 counties; Milton, 34 counties). Lines indicate the 25-, 50- and 100-mile distance bands. **b.** Mobility is grouped into six activity categories and decomposed into within-area (within-area flow comprises visits with both endpoints inside the focal unit. For the regional analysis, the focal unit is the entire affected region), inflow and outflow components relative to the affected region, A. **c.** Recovery estimation illustrated for within-region Retail & Leisure mobility during Hurricane Milton. Black points show observed daily trips; the blue line and shaded band show the calendar AR (1) counterfactual baseline and its 95% interval. The vertical dashed line marks landfall. The red point identifies the mobility trough, which is 28.7% below baseline. The green line illustrates the fitted Theil–Sen recovery trend, yielding an estimated recovery time of 5.2 days from landfall.

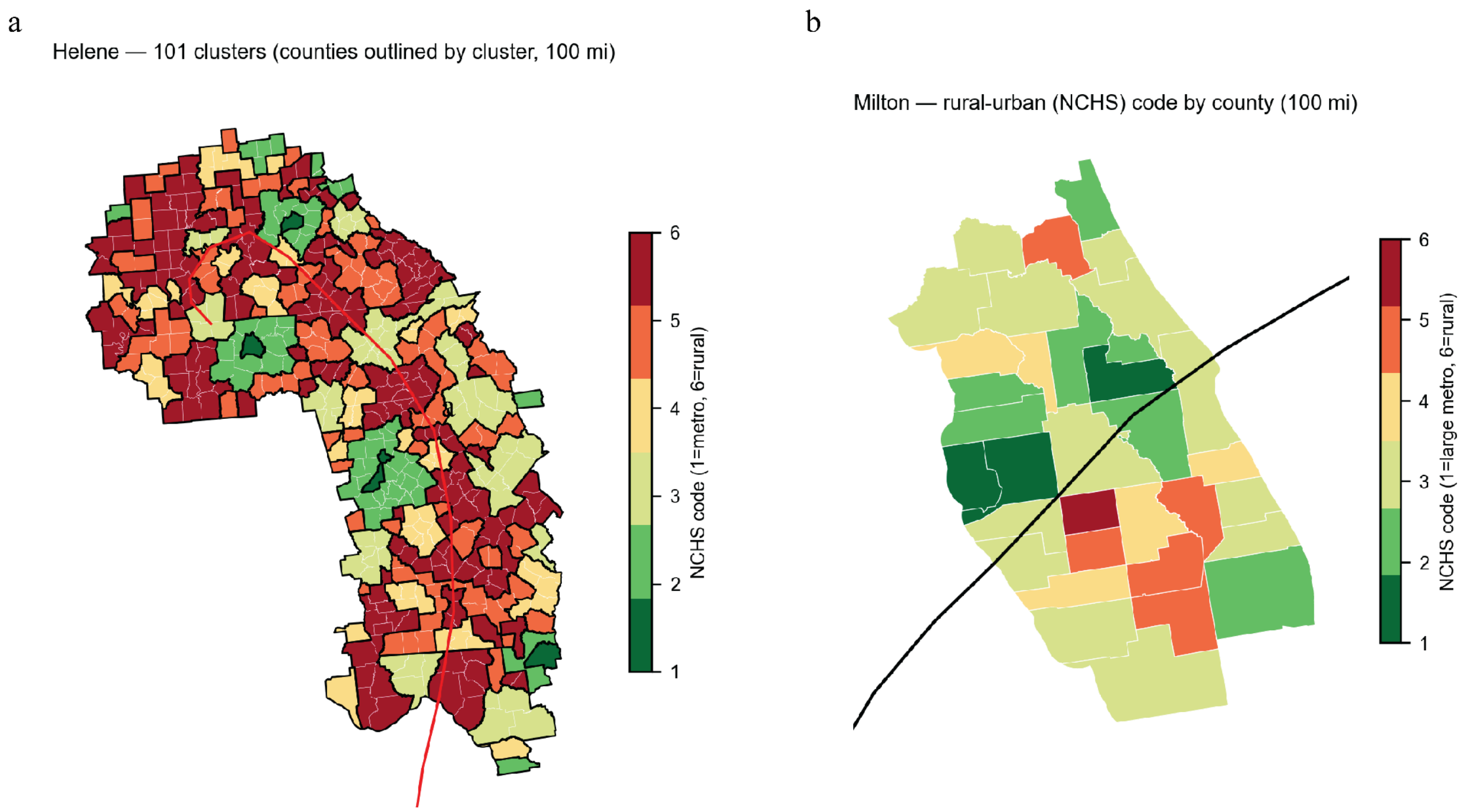

**Fig. S2. Local analysis units and urban–rural classification in the Helene and Milton study regions. a.** The 487 counties in the Helene study region are grouped into 101 spatially contiguous clusters whose constituent counties share the same 2023 National Center for Health Statistics (NCHS) urban–rural classification. Thick black outlines delineate clusters; thin white lines show county boundaries. **b.** The Milton study region comprises 34 counties, retained as individual analysis units. Both regions include counties whose centroids lie within 100 miles of the corresponding storm track. Colours indicate the six NCHS classes: 1, large central metropolitan; 2, large fringe metropolitan; 3, medium metropolitan; 4, small metropolitan; 5, micropolitan; and 6, noncore. Higher codes indicate more rural settings. The red line in **a** and black line in **b** show the respective storm tracks. Clustering preserves the urban–rural classification while aggregating Helene county mobility into larger local analysis units.

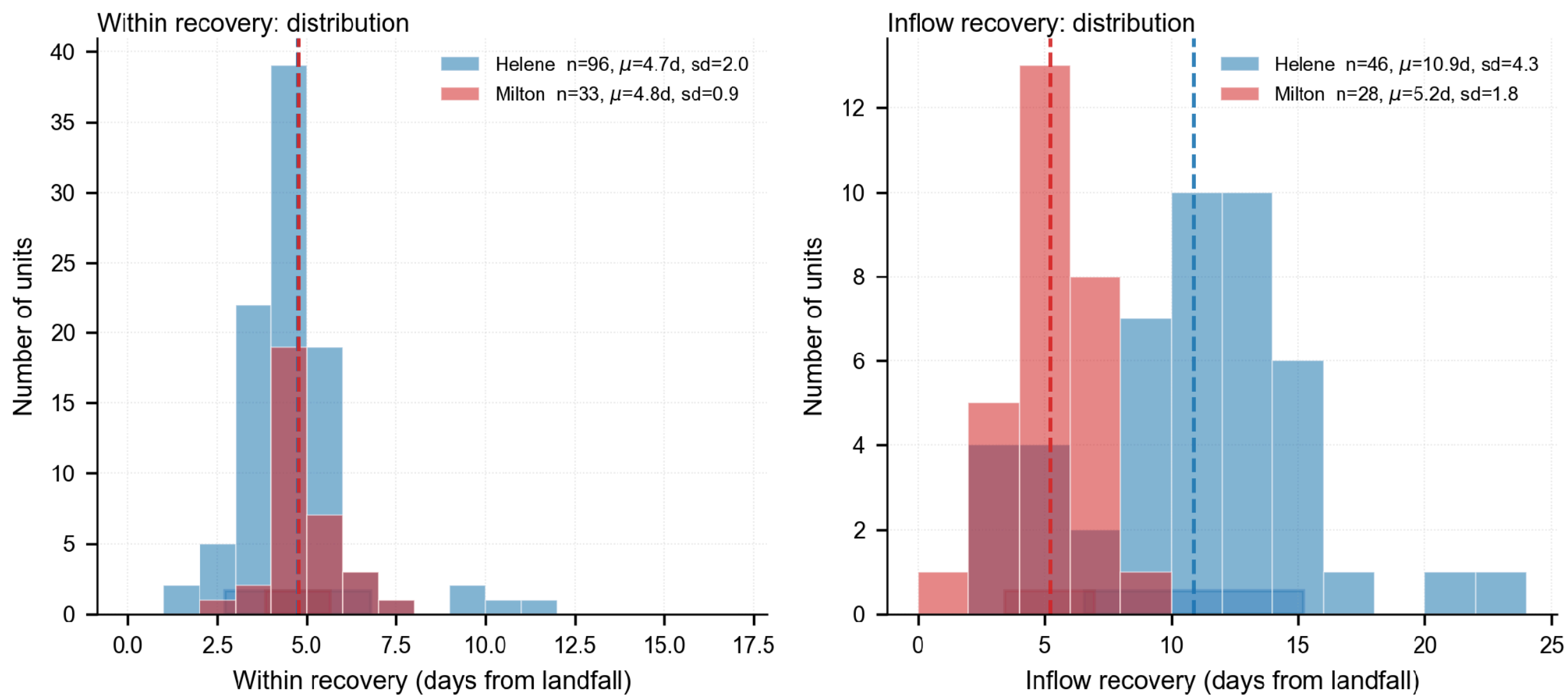

**Fig. S3 Distributions of local mobility recovery times after Hurricanes Helene and Milton.** Histograms show estimated recovery times for within-area mobility (left) and inflow (right), measured in days from landfall to the fitted return to counterfactual mobility. Blue denotes Helene clusters and red denotes Milton counties. Only units meeting the flow-specific volume threshold and having valid recovery estimates are included: within-area, 96 Helene clusters and 33 Milton counties; inflow, 46 clusters and 28 counties. Dashed vertical lines indicate means, and short horizontal bands along the baseline show one standard deviation on either side of the mean. Legends report sample size (n), mean (μ) and standard deviation (sd). Mean within-area recovery times were similar between storms (4.7 days for Helene and 4.8 days for Milton), whereas mean inflow recovery was longer after Helene (10.9 versus 5.2 days).

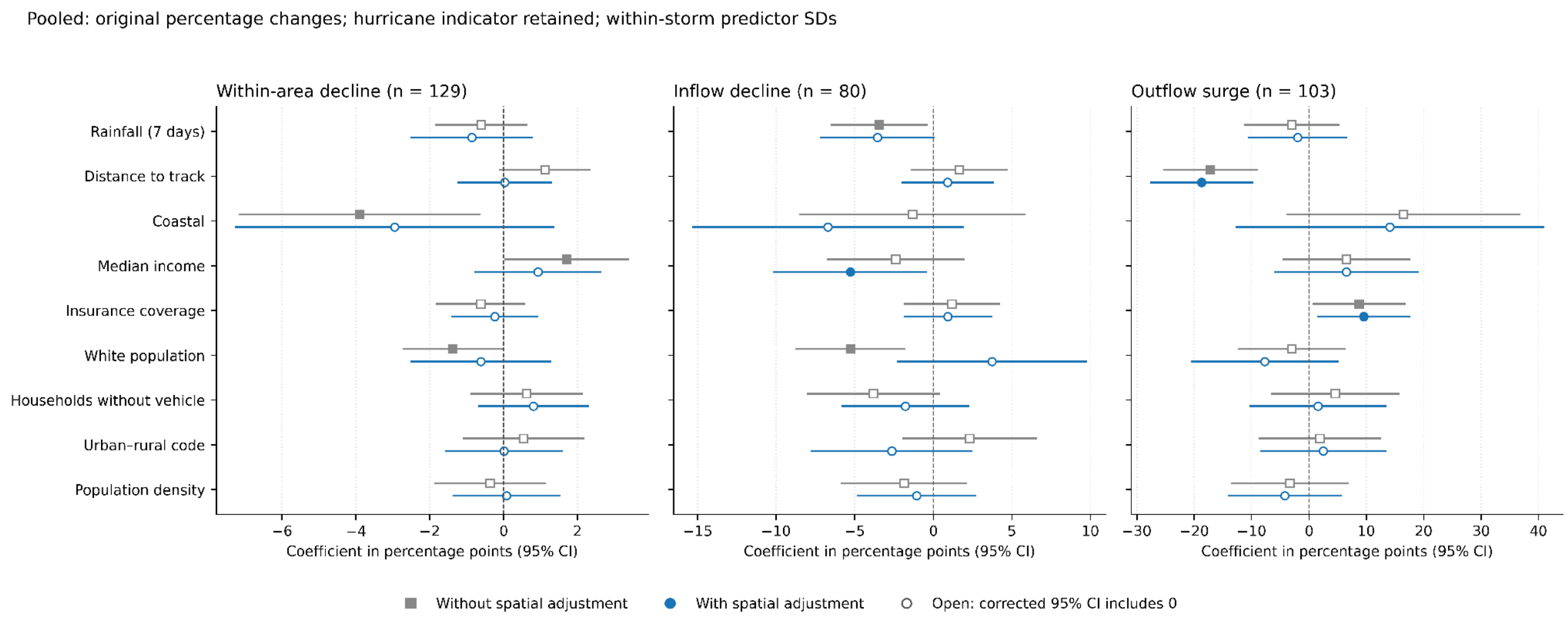

**Fig. S4 Pooled associations with mobility disruption after Helene and Milton.** Coefficient estimates for within-area decline (n = 129), inflow decline (n = 80) and outflow surge (n = 103) from nine-covariate models without spatial adjustment (grey squares) and with separate spatial smooths for each hurricane (blue circles). Both models retain a hurricane indicator. Outcomes are raw percentage changes; coefficients represent percentage-point differences per within-storm predictor standard deviation, except coastal location, which compares coastal with inland units. Bars indicate 95% confidence intervals, incorporating smoothing uncertainty for spatial models. Filled markers indicate intervals excluding zero; open markers indicate intervals including zero. Negative coefficients indicate deeper declines for within-area and inflow movement, or smaller outflow surges. The dashed line marks zero association.

**Table S1 Activity categories, NAICS sectors and POI counts**

| Activity category | NAICS sectors (codes) | Helene POI records | Milton POI records |
|---|---|---|---|
| Travel | 48–49 | 7,680 | 5,627 |
| Work & Professional | 11, 21, 23, 31–33, 42, 51–56, 81 | 217,809 | 148,980 |
| Health | 62 | 83,830 | 52,731 |
| Education | 61 | 13,108 | 6,633 |
| Retail & Leisure | 44–45, 71–72 | 113,818 | 66,021 |
| Urban Government | 92 | 6,045 | 2,609 |
| **Total across six categories** | | **442,290** | **282,601** |

NAICS, North American Industry Classification System; POI, point of interest. Sector codes summarize the original NAICS lookup used to assign records to the six activity groups; six-digit codes were matched using their first four digits, and available two- or three-digit codes were matched directly. Counts are POI records in the weekly extract for 9–15 September 2024, a common reference week before both storms, with destinations in the 100-mile study regions (487 Helene counties; 34 Milton counties), before local-unit visit-volume restrictions. Each retained weekly record contributes one count; records missing required input fields or a mapped NAICS code are excluded. The extract lacks establishment identifiers, so these are weekly record counts, not counts of unique POIs across the full study period. Utilities (NAICS 22) are excluded.

**Table S2 Local analysis samples.**

| Sample or outcome | Helene | Milton |
|---|---:|---:|
| Underlying counties | 487 | 34 |
| Primary local units | 101 clusters | 34 counties |
| Within-area volume retained | 96 | 33 |
| Inflow volume retained | 52 | 28 |
| Outflow volume retained | 74 | 29 |
| Valid within-area recovery after volume restriction | 96 | 33 |
| Valid inflow recovery after volume restriction | 46 | 28 |

Rows refer to different eligibility criteria, not successive exclusions across flow types. A unit can be retained for one flow and excluded for another. Sample sizes reflect two eligibility criteria: the flow-specific volume threshold described in the main Methods and, for recovery analyses, availability of a valid recovery estimate. The volume threshold uses mean observed daily visits during the forecast period, including the storm, so eligibility can be influenced by storm-related mobility changes. An area may qualify for one flow but not another. Undefined recovery times are excluded from recovery summaries rather than assigned zero days.

**Table S3 Local disruption and recovery**

| Outcome | Helene median [IQR] | Milton median [IQR] | P |
|---|---|---|---|
| Within recovery (days) | 4.4 [3.7, 5.1] | 4.5 [4.1, 5.4] | 0.294 |
| Inflow recovery (days) | 11.1 [8.7, 13.4] | 5.1 [4.2, 6.2] | <0.001 |
| Within peak deviation (%) | −15 [−17, −11] | −27 [−33, −20] | <0.001 |
| Inflow peak deviation (%) | −20 [−29, −13] | −33 [−42, −24] | <0.001 |
| Outflow maximum deviation (%) | 19 [5, 31] | 18 [−3, 55] | 0.991 |

IQR denotes the 25th–75th percentile interval. Sample sizes are the corresponding retained or valid-recovery counts in Table S2. P values are from two-sided Mann–Whitney comparisons of local units. They are descriptive comparisons of distributions, without adjustment for spatial dependence or multiple testing; a nonsignificant result does not establish equivalence. Values are rounded as in the frozen analysis summary. [1]

The median inflow contrast is 6.0 days, or approximately 2.2 times as long under Helene, calculated from the rounded medians. The corresponding within-area contrast is −0.1 day. Milton has the deeper within-area and inflow deficits, while the outflow maxima have similar medians and substantially different interquartile spans. The inflow recovery gap is therefore not a general ordering of recovery by storm or by disruption depth.

The canonical 10-day trough search yields 89 valid Helene inflow estimates before volume restriction, compared with 72 in the earlier 7-day implementation. A shorter search can miss delayed troughs and alter recovery eligibility. The peak disruption windows are independent of this recovery search, so changing the trough window does not change the peak-deviation outcomes. [1]

## Table S4 Predictor sources and distributions by storm

| Predictor and units | Data source | Helene<br>101 clusters | Milton<br>34 counties |
|---|---|---|---|
| Seven-day precipitation (mm) | PRISM daily 4-km precipitation grids, 2024 [5] | 107.7 [76.6, 131.5] | 148.3 [77.5, 231.2] |
| Distance to storm track (miles) | 2024 storm-track shapefiles; TIGER/Line 2023 county centroids [1,8] | 69.7 [41.5, 95.3] | 54.3 [31.2, 84.1] |
| Coastal status, yes | TIGER/Line 2023 counties; Natural Earth 1:10 million coastline [8,9] | 12 (11.9%) | 19 (55.9%) |
| Median household income (US$) | ACS 2018–2022 five-year estimates; B19013 [4] | 56,044 [50,022, 61,955] | 65,233 [54,076, 71,134] |
| White population (%) | ACS 2018–2022 five-year estimates [4] | 83.7 [67.5, 90.6] | 80.1 [74.5, 84.4] |
| Insurance coverage (%) | ACS 2018–2022 five-year estimates; B27010 [4] | 98.8 [98.4, 99.1] | 98.5 [98.3, 98.7] |
| Households without a vehicle (%) | ACS 2018–2022 five-year estimates; B08201 [4] | 5.7 [4.5, 6.6] | 4.7 [4.4, 5.2] |
| Population density (residents/mi²) | ACS 2018–2022 population; TIGER/Line 2023 land area [4,8] | 95.2 [65.2, 182.9] | 319.9 [202.0, 597.6] |
| NCHS urban–rural classification | 2023 NCHS six-level county classification [2] | n (%) | n (%) |
| 1 Large central metro | | 1 (1.0%) | 3 (8.8%) |
| 2 Large fringe metro | | 10 (9.9%) | 7 (20.6%) |
| 3 Medium metro | | 11 (10.9%) | 13 (38.2%) |
| 4 Small metro | | 20 (19.8%) | 5 (14.7%) |
| 5 Micropolitan | | 36 (35.6%) | 5 (14.7%) |
| 6 Noncore | | 23 (22.8%) | 1 (2.9%) |

Values are median [25th, 75th percentile] for continuous variables and n (%) for coastal status and NCHS classes. Each cluster or county contributes one observation; summaries are unweighted across units and cover all 135 primary local units before flow-specific volume restrictions (Table S2). All nine predictors are complete. Coastal counties intersect a 100-m buffer around the coastline. The eight numeric predictors, including NCHS code, were standardized using the pooled 135-unit means and standard deviations before model-specific filtering; coastal status remained 0/1. Table values are on the original scales. ACS, American Community Survey; NCHS, National Center for Health Statistics. [1]

**Table S5 Helene associations with 95% HDIs excluding zero**

| Outcome and predictor | Posterior mean | 95% HDI | n |
|---|---|---|---|
| Within deficit: coastal | −4.9 | −8.6 to −1.3 | 96 |
| Inflow deficit: precipitation | −9.0 | −14.3 to −3.5 | 52 |
| Inflow deficit: percentage White | −7.0 | −11.0 to −3.4 | 52 |
| Outflow maximum: track distance | −19.9 | −29.6 to −9.2 | 74 |

HDI, highest-density interval. Negative coefficients on within-area or inflow deviations indicate deeper deficits; negative coefficients on outflow maxima indicate a smaller maximum. This table reports the four selected associations, not a complete coefficient inventory. None of Milton's 27 predictor–outcome intervals excludes zero. That finding is consistent with substantial uncertainty at $n = 28–33$ and cannot establish an absence of associations. The intervals have not been presented as a familywise multiplicity adjustment. [1]

Within Helene, inflow disruption has Spearman $\rho = 0.11$ with track distance ($P = 0.431$, $n = 52$). Inflow recovery has $\rho = 0.07$ with NCHS code ($P = 0.637$, $n = 46$) and $\rho = -0.46$ with track distance ($P = 0.001$). Thus the cross-storm recovery contrast does not demonstrate a within-region rural–urban gradient or identify a causal effect of rurality.

**Table S6 Helene inflow coefficients before and after spatial adjustment**

| Predictor | Without smooth β (P) | With smooth β (P) |
|---|---|---|
| Percentage White | −4.00 (0.046) | 0.37 (0.891) |
| Coastal status | −0.10 (0.984) | −13.36 (0.040) |

The percentage-White association attenuates after spatial adjustment, whereas the coastal coefficient becomes more negative. This comparison shows sensitivity to the spatial specification; it does not prove that socioeconomic composition has no effect. Smooth geography and geographically patterned covariates can share explanatory variation. The GAM coefficients should not be substituted for the Bayesian coefficients in Table S6.

Residual Moran's I under Queen contiguity is at most 0.07 with $P > 0.20$ in five of the six storm–flow models. Milton inflow is the exception ($I = 0.22$, $P = 0.039$). Raw-outcome Moran's I ranges from 0.11 to 0.35. Residual dependence in Milton inflow limits interpretation of a spatially independent regression for that outcome. [1,7]

## Supplementary References

1. Qing1011. milton_helene_mobility_category_analysis. Branch phase1-manuscript-numbers, commit 612510512ca3552cda84083b0ca780ae7471609c. Frozen summary: results/npj_100mi/MANUSCRIPT_NUMBERS_README.md; methods and sensitivity: notes/methods_baseline_model.md and notes/findings.md; implementation: npj/notebook/. https://github.com/Qing1011/milton_helene_mobility_category_analysis/tree/612510512ca3552cda84083b0ca780ae7471609c

2. National Center for Health Statistics. NCHS Urban–Rural Classification Scheme for Counties, 2023 scheme and documentation. https://www.cdc.gov/nchs/data-analysis-tools/urban-rural.html

3. Sen, P. K. Estimates of the regression coefficient based on Kendall's tau. Journal of the American Statistical Association 63, 1379–1389 (1968). https://doi.org/10.1080/01621459.1968.10480934

4. U.S. Census Bureau. American Community Survey 2022 five-year estimates. https://api.census.gov/data/2022/acs/acs5.html

5. PRISM Climate Group, Oregon State University. PRISM time series datasets. https://prism.oregonstate.edu/data/

6. Wood, S. N. Generalized Additive Models: An Introduction with R, second edition (CRC Press, 2017). https://doi.org/10.1201/9781315370279

7. Moran, P. A. P. Notes on continuous stochastic phenomena. Biometrika 37, 17–23 (1950). https://doi.org/10.1093/biomet/37.1-2.17

8. U.S. Census Bureau. TIGER/Line Shapefiles, 2023 county boundaries and land-area attributes. https://www.census.gov/geographies/mapping-files/time-series/geo/tiger-line-file.2023.html

9. Natural Earth. Coastline, 1:10 million scale. https://www.naturalearthdata.com/downloads/10m-physical-vectors/10m-coastline/